\documentclass[10pt,conference]{IEEEtran}
\IEEEoverridecommandlockouts
\usepackage{enumitem}
\usepackage{arydshln}
\usepackage{subcaption}
\usepackage{amsmath}
\usepackage[hyphens]{url}
\usepackage{color}
\usepackage{booktabs}
\usepackage{multirow}
\usepackage{tikz}
\usetikzlibrary{shapes.geometric,arrows,decorations.markings, arrows.meta}
\usetikzlibrary{calc}

\usepackage{cite}
\usepackage{amsmath,amssymb,amsfonts}
\usepackage{algorithmic}
\usepackage{graphicx}
\usepackage{textcomp}
\usepackage{xcolor}
\usepackage[official]{eurosym}
\usepackage{hyperref}
\usepackage{tcolorbox}

\newif\ifdraft
 \draftfalse %% turn off comments

\ifdraft
    \newcommand\ns[1]{{\color{purple}[NS]: #1}}
    \newcommand\og[1]{{\color{green!60!black}[OG]: #1}}
\else
    \newcommand\ns[1]{}
    \newcommand\og[1]{}
\fi

\def\BibTeX{{\rm B\kern-.05em{\sc i\kern-.025em b}\kern-.08em
    T\kern-.1667em\lower.7ex\hbox{E}\kern-.125emX}}
\begin{document}

\title{Empirical assessment of the perception of graphical threat model acceptability
% {\footnotesize \textsuperscript{*}Note: Sub-titles are not captured for https://ieeexplore.ieee.org  and
}

\author{
\IEEEauthorblockN{
% 1\textsuperscript{st} 
Nathan Daniel Schiele}
\IEEEauthorblockA{
\textit{LIACS, Leiden University}\\
Leiden, Netherlands \\
0000-0003-1186-1503}
\and
\IEEEauthorblockN{
% 2\textsuperscript{nd} 
Olga Gadyatskaya}
\IEEEauthorblockA{
\textit{LIACS, Leiden University}\\
Leiden, Netherlands \\
0000-0002-3760-9165}
}

% \author{
% \IEEEauthorblockN{Anonymized for submission}
% }

\maketitle

\pagestyle{plain}

\begin{abstract}

Threat modeling (TM) is an important aspect of risk analysis and secure software engineering. Graphical threat models are a recommended tool to analyze and communicate threat information. However, the comparison of different graphical threat models, and the acceptability of these threat models for an audience with a limited technical background, is not well understood, despite these users making up a sizable portion of the cybersecurity industry.

We seek to compare the acceptability of three general, graphical threat models, Attack-Defense Trees (ADTs), Attack Graphs (AGs), and CORAS, for users with a limited technical background. We conducted a laboratory study with 38 bachelor students who completed tasks with the three threat models across three different scenarios assigned using a Latin square design. Threat model submissions were qualitatively analyzed, and participants filled out a perception questionnaire based on the Method Evaluation Model (MEM). 

We find that both ADTs and CORAS are broadly acceptable for a wide range of scenarios, and both could be applied successfully by users with a limited technical background; further, we also find that the lack of a specific tool for AGs may have impacted the perceived usefulness of AGs. 
We can recommend that users with a limited technical background use ADTs or CORAS as a general graphical TM method. Further research on the acceptability of AGs to such an audience and the effect of a dedicated TM tool support is needed.

\end{abstract}

\maketitle              % typeset the header of the contribution

\newcommand{\etal}{et al.}
\newcommand{\id}[2]{#1-#2}
\newcommand{\hypothesis}[1]{$\text{H}_\text{#1}$}
\newcommand{\RQ}[1]{\textbf{RQ#1}}

\newcommand{\ICS}{NCS}
\newcommand{\SEC}{CS}

\newcommand{\AND}{AND}
\newcommand{\SAND}{SAND}
\newcommand{\OR}    {OR}

\newcommand{\APAlong}{absolute perception of acceptability}
\newcommand{\CPAlong}{comparative perception of acceptability}
\newcommand{\APA}{APA}
\newcommand{\CPA}{CPA}

\newcommand{\rank}[1]{\texttt{R#1}}

\newcommand{\hResponse}[1]{\texttt{#1} - }

\newcommand{\anonfoot}{\footnote{Anonymized for submission}}

\newcommand{\qIndent}{4em}
\newcommand{\qsIndent}{2em}
\newcommand{\surveyq}[1]{\textbf{#1:}}

\newcommand{\highlight}[1]{
\begin{tcolorbox}[
    colback=yellow!20, 
    boxrule=0.4pt, 
    arc=1pt,
    left=0mm, 
    right=0mm, 
    top=0mm, 
    bottom=0mm,
]
    #1
\end{tcolorbox}
}
\section{Introduction}

The ever-increasing number of threats requires better processes and tools to ensure sufficient security outcomes. \emph{Threat modeling} (TM) is a recommended practice for risk analysis~\cite{andersonSecurityEngineeringGuide2020} and security-by-design~\cite{shostackThreatModelingDesigning2014}. In the context of secure software engineering, threat modeling is the process of identifying and documenting potential threats to a software system and the potential consequences of those threats~\cite{yamamoto2025cottage}.  Threat modeling helps to develop secure software~\cite{yskout2020threat,howard2003inside,apvrille2005secure} by identifing potential vulnerabilities in a system early in the development process and to proposing suitable countermeasures, thus enhancing security outcomes~\cite{shostackThreatModelingDesigning2014}. The process of threat modeling can occur at all levels of an organization~\cite{shostackThreatModelingDesigning2014,schneierSecretsLiesDigital2000,andersonSecurityEngineeringGuide2020} and with a diverse set of stakeholders, including those with a limited technical (computer science) background~\cite{baldwinInformationseekingBehaviorSecurities1997,steinbartInfluenceGoodRelationship2018,dev2023models}. Despite the value that threat modeling can provide, the application of TM in practice is less clear~\cite{verreydt2024threat, schiele2024nobody}. By better understanding the strengths of different TM methods (in terms of application to different scenarios and audiences), we can better understand how to apply threat modeling.

\emph{Graphical threat models} are representations of attacks and defenses that are presented visually and defined in a graphical structure. One well-known example of a graphical threat model is the \emph{Attack Tree} (AT)~\cite{schneierAttackTrees1999} or its derivative form, and one of the models used in our study, the \emph{Attack-Defense Tree} (ADT)~\cite{kordyFoundationsAttackDefense2011}. This threat models is drawn from the perspective of an attacker and represent hierarchically different components in an attack scenario. Another graphical threat model called \emph{Attack Graphs} (AGs)~\cite{sheynerAutomatedGenerationAnalysis2002} represents similar information but in a state-based model. Finally, \emph{CORAS}~\cite{solhaugCORASLanguageWhy2014} is a graphical threat model that explicitly includes different aspects of an attack scenario, such as a threat source of a vulnerability. These models are evaluated in this work.

The Method Evaluation Model (MEM)~\cite{moodyMethodEvaluationModel2003} is a commonly used methodology for threat model evaluation studies~\cite{labunetsExperimentalComparisonTwo2013, labunetsExperimentComparingTextual2014,schieleAcceptability2025,borstler2024acceptance}. The MEM focuses on ``\emph{acceptability}'', which is a quality measurement of how acceptable a method is for a given task for a specific audience. The MEM describes the effect of actual and perceived performance and usability of a method on the intention to use and the actual use of that method. In our work, we use the MEM to compare the acceptability of different graphical threat models.

Specifically, we conducted a study ($n=38$) to understand the acceptability of three aforementioned general graphical threat models: Attack-Defense Trees (ADTs)~\cite{kordyFoundationsAttackDefense2011}, Attack Graphs (AGs)\cite{sheynerAutomatedGenerationAnalysis2002}, and CORAS~\cite{solhaugCORASLanguageWhy2014}, for an audience with a very limited technical (computer science) background. We aim to understand how the acceptability of these models differs between the models themselves and how the acceptability of these models may differ between different scenarios.  We find that the acceptability of the models does not differ significantly, and that the scenario the model was used on does not affect the acceptability of the model. We also find that the lack of a specific tool for AGs may have impacted the perceived usefulness of AGs. 
We further find that both ADTs and CORAS are broadly acceptable for a wide range of scenarios, and both could be applied successfully by users with a limited technical background.

\subsection{Data Availability}
\label{sec:data}

The complete set of anonymized participant responses to survey questions, including responses to questions that were not used in this work, is available as a \texttt{.csv} file. The questionnaire, qualitative evaluation rubric, and other study materials are also available. These data are available at the following Zenodo repository: \url{https://doi.org/10.5281/zenodo.15269163}.

\section{Research Objectives}
\label{sec:research}

Our research question for this study is as follows,

\textbf{RQ:} \emph{How does the acceptability of general graphical threat models differ for participants with a limited technical background?}

To answer this research question, we have developed the following hypotheses:

\begin{itemize}
    \item \textbf{\hypothesis{1}}: There will be a significant difference in the absolute acceptability between the threat models.
    \item \textbf{\hypothesis{2}}: There will be a significant difference in the comparative perception of acceptability of the different threat models.
    \item  \textbf{\hypothesis{3}}:  The absolute acceptability of the threat models will differ between different scenarios.
    
\end{itemize}

These hypotheses contain two terms, \emph{absolute acceptability} (AA) and \emph{comparative perception of acceptability} (CPA).  \emph{Absolute acceptability} refers to the acceptability of a threat model in isolation; it is a property of the threat model itself. We will measure the components of acceptability (per the MEM) for each threat model (and for each threat model per scenario), and compare these components to each other to determine any difference in absolute acceptability. \emph{Comparative perception of acceptability} in contrast refers to the perception of acceptability of a threat model compared to other threat models. We ask participants to rank the threat models according to components of acceptability and compare these rankings to each other to determine differences in CPA.

To offer further clarification, it may be the case that all models are perceived as acceptable for the tasks assigned, so the absolute acceptability of the models would not differ. However, when having participants rank the models, it may be the case that one model is ranked higher than the others, so the comparative perception of acceptability of the models could differ (while absolute acceptability does not). We can measure actual usage for absolute acceptability, but for rankings, we can only measure perception -- hence, the distinction between AA and comparative \emph{perception} of acceptability.

To test these hypotheses, we conducted an experiment with student participants with a limited technical background ($n = 38$). Students were given an assignment to create threat models based on news articles about cyber attacks. The students were then asked to fill out a questionnaire about their perceptions of the threat models. The questionnaire was based on the Method Evaluation Model (MEM)~\cite{moodyMethodEvaluationModel2003}. The MEM is an extension of the Technology Acceptance Model (TAM)~\cite{borstler2024acceptance,davisPerceivedUsefulnessPerceived1989}. It is a popular method~\cite{borstler2024acceptance}, which has been used in several previous threat model acceptability studies~\cite{labunetsExperimentalComparisonTwo2013, labunets2017model,labunets_no_2018,broccia_assessing_2024,broccia2025evaluating,schieleAcceptability2025}; with these studies also including examinations of both CORAS and ADTs, albeit with a different focus. The MEM is a model that describes the effect of actual and perceived performance and usability of a method on the intention and actual use of that method. Figure~\ref{fig:mem} shows the MEM.

Following from the results of Labunets~\etal\ in \cite{labunets2017model}, who found that graphical and tabular threat modeling methods differ in performance with respect to different tasks, we seek to understand if such a difference could exist within different graphical threat modeling methods. Instead of assessing the threat modeling methods on their task performance, we focus on the overall acceptability of these models. We examine general acceptability as this would better examine the potential performance of a model in a general (industry) context, as practitioners do not necessarily use models as they are intended~\cite{kaurThreatModelingVery2025,verreydt2024threat}. Further, we are specifically interested in studying the acceptability of these models for participants with a limited technical background, as threat modeling is an activity that involves diverse stakeholders, some of whom may lack a computer science background~\cite{verreydt2024threat} and thus can be potentially disadvantaged by the choice of the TM method.

\section{Background}
\label{sec:background}

There exist many methods and notations for threat modeling~\cite{shevchenko2018threat,xiong2019threat,tuma2018threat,hongSurveyUsabilityPractical2017}. In our work, we examine attack graphs, attack-defense trees, and CORAS. When it is non-ambiguous from the context, we refer to them as \emph{models} (meaning a \emph{modeling notation} or TM \emph{method}). 

\emph{Attack graphs} (AGs) are a state-based graphical threat model~\cite{sheynerAutomatedGenerationAnalysis2002,lallieReviewAttackGraph2020}. Every node in an AG represents the current state of a system, while edges between nodes represent transitions between states.  The nodes are labeled with the state of the system, and the edges are labeled with the actions that can be taken to transition between states. States are indicated to either be unsafe, where the system results in an unwanted state, or safe, where the system results in an acceptable state. Defenses to attacks can be modeled by including safe states in the model. 

\emph{Attack-defense trees} (ADTs) are a component-based graphical threat model~\cite{schneierAttackTrees1999,mauwFoundationsAttackTrees2006,kordyFoundationsAttackDefense2011,lallieReviewAttackGraph2020}. An ADT is a directed, rooted tree, where every node represents either a goal or a component of an attack or defense. The root node represents the overall goal of an attacker (or a defender). Child nodes in ADTs are either in an \AND\ relationship, where all children must be satisfied for the parent to be satisfied, or an \OR\ relationship, where only one child must be satisfied for the parent to be satisfied~\cite{schneierAttackTrees1999,mauwFoundationsAttackTrees2006}. Nodes can have one countermeasure, which is the opposite type (defense or attack), and this edge is typically represented as a dashed line~\cite{kordyFoundationsAttackDefense2011}. Security controls are modeled using defense nodes and countermeasure edges.

\emph{CORAS} is an aspect-based graphical threat model~\cite{solhaugCORASLanguageWhy2014}. CORAS diagrams are composed of different aspects of attacks. The aspects are the risk source, vulnerabilities, threat scenario, treatment, incident, and asset. These aspects are represented by nodes in the graph and differentiated by defined symbols. Edges between nodes represent relationships between these aspects. Defenses can be modeled by including treatments to threat scenarios~\cite{solhaugCORASLanguageWhy2014}.

 These three models were selected as all are graphical and previous research has shown that graphical models are better suited than textual models for some tasks~\cite{labunetsExperimentComparingTextual2014}. Additionally, these three models are all general threat models, meaning they can be applied to a wide range of scenarios, including non-cybersecurity-related scenarios. This is in contrast to other methods, such as STRIDE~\cite{scandariato2015descriptive} which is a system specific TM method. Finally, we selected these three models as they all represent attack information differently. AGs present information based on states and transitions between states, ADTs present information based on components and relationships between components, and CORAS presents information based on aspects of a scenario. This distinction allows us to understand how the different representations of the same information affect the acceptability of the models. These models are popular in the literature~\cite{labunetsExperimentComparingTextual2014,opdahlExperimentalComparisonAttack2009,broccia2025evaluating,schieleAcceptability2025, haqueEvolutionaryApproachAttack, de_la_vara_empirical_2020, munoz-gonzalezEfficientAttackGraph2017, sommestadEmpiricalTestAccuracy2015,stergiopoulos2022automatic,yamamoto2025cottage, schiele2021novel}, and are recommended for use in industry~\cite{shostackThreatModelingDesigning2014,schneierSecretsLiesDigital2000,andersonSecurityEngineeringGuide2020,lund2010model,tarandach2020threat,reversinglabs2024attacktrees}. While methods like STRIDE might be more commonly used for TM in software engineering, general TM notations are also widely applied. For example, ADTs are recommended by industry resources~\cite{shostackThreatModelingDesigning2014,tarandach2020threat,reversinglabs2024attacktrees} to complement STRIDE for in-depth analysis of critical threats.

Our study is motivated by the need to make threat modeling more accessible, especially for practitioners with a limited technical background. Verreydt et al.~\cite{verreydt2024threat} have shown that, while the roles involved in software development (developers, architects, product owners, and the security team) were central to conducting the TM process, the outcomes are often communicated to managers and information security officers. Moreover, the perception that TM requires strong technical background is one of the reasons why management is not involved directly~\cite{verreydt2024threat}, while direct involvement by management is considered a best practice~\cite{ingalsbe2008threat}. Thus, we study a population with a limited technical background (which can be reasonably expected in a software development company) to understand how acceptable the chosen TM methods are for this group.

\section{Related Work}
\label{sec:related-work}

Threat model acceptability, efficacy, and comprehensibility were examined in the literature. These studies all vary in their focus. 
To our knowledge, we are the first to \emph{compare the acceptability of three different general graphical threat models (ADTs, AGs, and CORAS) for participants with a limited technical background}.

% \subsection{Threat model acceptability studies}

The work strongly related to ours is a study by Labunets~\etal; a study conducted with 28 master's students with a computer science background (12 with a background in security) using the MEM to compare the acceptability of CORAS to  Security Requirements Engineering Process (SREP), a textual threat model~\cite{labunetsExperimentalComparisonTwo2013}. They found participants had a preference for the graphical threat model over the textual model in their perception, but that both models had different strengths in their actual use. We build on this work by comparing the acceptability of three graphical threat models to see if there is a preference for a specific type of graphical threat model.  Further, our study focuses on participants with a limited technical background to see if the results are generalizable to a wider audience.

Several studies have used a similar between-groups design to compare multiple methods by having participants perform the same tasks using several threat modeling methods. Opdahl and Sindre compared attack trees to misuse cases, finding that attack trees were more effective, but the participants had similar perceptions of the two techniques~\cite{opdahlExperimentalComparisonAttack2009}. In a very relevant work, Lallie~\etal\ compared attack graphs to fault trees (considered as a variant of attack trees), finding attack graphs more effective~\cite{lallieEmpiricalEvaluationEffectiveness2017}, especially for participants with a computer science background. Katta~\etal\ compared the performance, understanding, and perception of misuse sequence diagrams and misuse case maps in an experiment with student participants, finding that the models perform similarly~\cite{kattaComparingTwoTechniques2010}. De La Vara~\etal\ compared a Systems Process Engineering Metamodel-like diagrams to text descriptions; they found that the model was statistically significantly more effective than the text descriptions as a means of analysis~\cite{de_la_vara_empirical_2020}. Finally, Diallo~\etal\ compared Common Criteria, misuse cases, and attack trees, finding advantages and disadvantages for each approach~\cite{dialloComparativeEvaluationThree2006}.

There have been several empirical studies focusing on the acceptability of a single graphical threat modeling method. Broccia~\etal\ examined attack tree acceptability using a study design with MEM and 25 human subjects~\cite{broccia_assessing_2024}. This study was also recently replicated in another experiment with 49 subjects~\cite{broccia2025evaluating}. Both Broccia~\etal\ studies found that attack trees had a sufficient level of acceptability. In our previous work, we conducted a MEM-based study with 102 participants, comparing two different populations, highly technical and limited technical, examining the acceptability of attack-defense trees~\cite{schieleAcceptability2025}. We found that the acceptability of the models was largely consistent across the two populations. 
% However, the study~\cite{schieleAcceptability2025} only examined ADTs and did not compare them with other methods. 
Katsikeas et al.~\cite{katsikeasEmpiricalEvaluationThreat2024} performed an empirical study comparing attack graphs generated by domain experts to those created by computers, finding that the computer-generated AGs were more effective.

Finally, several empirical studies on non-graphical threat modeling methods have been conducted. For example, Wuyts et al.~\cite{wuytsEmpiricalEvaluationPrivacyfocused2014} empirically evaluated the LINDDUN method in a series of 3 studies. Bernsmed et al.~\cite{bernsmedAdoptingThreatModelling2022}, Scandariato et al.~\cite{scandariato2015descriptive}, and Tuma and Scandariato~\cite{tuma2018two} all evaluated the acceptability of STRIDE~\cite{kohnfelder1999threats} with students as participants.

\section{Research Method}
\label{sec:methods}

In this section, we outline our process for conducting the study. We discuss the theoretical framework of the study, the study design, the participants and training, data processing and analysis, and the ethical considerations of our study.

% \subsection{Methodology}
% \label{ssec:methods-theory}

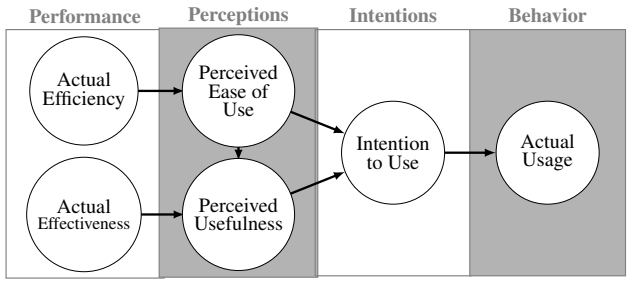
\begin{figure}
    \resizebox*{\columnwidth}{!}{\tikzstyle{memnode} = [circle,
minimum size=2.5cm,
text centered,
draw=black,
fill=white]

\tikzstyle{memnodeinvis} = [circle,
minimum size=2.5cm,
text centered,
draw=white,
text=white]

\definecolor{icscolor}{RGB}{31,119,180}
\definecolor{seccolor}{RGB}{255,127,14}
\definecolor{thirdcolor}{RGB}{44,160,44}

% \tikzstyle{ics} = [rectangle, rounded corners,
% minimum width=1cm,
% minimum height=1cm, text centered,
% draw=black, fill=icscolor]

% \tikzstyle{sec} = [rectangle, rounded corners,
% minimum width=1cm,
% minimum height=1cm, text centered,
% draw=black, fill=seccolor]

% \tikzstyle{study} = [rectangle,
% rounded corners,
% minimum width=1cm,
% minimum height=1cm,
% text centered,
% draw=black,
% fill=thirdcolor]

% \tikzstyle{decision} = [diamond,
% minimum width=3cm,
% minimum height=1cm,
% text centered,
% draw=black,
% fill=green!30]
\tikzstyle{arrow} = [ultra thick,-latex]

\newcommand{\NodeTextSize}{\Large}
\newcommand{\smallerNodeText}{\large}

\begin{tikzpicture}[node distance=3cm,]

\node (AE) [memnodeinvis, xshift=-3cm] {\NodeTextSize\shortstack{Actual\\Efficiency}};
\node (AEff) [memnodeinvis, below of=AE] {\NodeTextSize\shortstack{Actual\\{\smallerNodeText Effectiveness}}};%\\\hypothesis{1}, \hypothesis{1}, \hypothesis{1}} };
\node (PEOU) [memnodeinvis, right of=AE, xshift=.75cm] {\NodeTextSize\shortstack{Percieved\\Ease of\\Use}};%\\\hypothesis{1}, \hypothesis{1}} };
\node (PU) [memnodeinvis, below of=PEOU] {\NodeTextSize\shortstack{Percieved\\Usefulness}};%\\\hypothesis{1}, \hypothesis{1}} };
\node (IU) [memnodeinvis, right of=PEOU, yshift=-1.5cm, xshift=.75cm] {\NodeTextSize\shortstack{Intention\\to Use}};%\\\hypothesis{1} }};
\node (AU) [memnodeinvis, right of=IU, xshift=.75cm ] {\NodeTextSize\shortstack{Actual\\Usage} };
    %     \node (sec1) [sec, below of=ics1] {\shortstack{\SEC\\students} };
    %     \node (ics2) [ics, right of=ics1] {\shortstack{\ICS\\lecture} };
    %     \node (sec2) [sec, right of=sec1] {\shortstack{\SEC\\lecture} };
    %     \node (ss) [study, right of=ics2, xshift=0.8cm] {\shortstack{Small\\study} };
    %     \node (ls) [study, right of=sec2, xshift=0.8cm] {\shortstack{Large\\study} };
    % \node (data) [study, right of=ss, xshift=0.3cm, yshift=-1cm] {\shortstack{Data\\collection} };

    % \draw [arrow] (sec1) -- (sec2);
    % \draw [arrow] (ics2) -- node[anchor=south] {Optional} (ss);
    % \draw [arrow] (ics2) -- (ls);
    % \draw [arrow] (sec2) -- (ss);
    % \draw [arrow] (sec2) -- node[anchor=north] {Mandatory} (ls);
    % \draw [arrow] (ss) -- node[anchor=south,pos=0.45, rotate=-25] {Consent} (data);
    % \draw [arrow] (ls) -- node[anchor=north,pos=0.45, rotate=25] {Consent} (data);
    \Large
\draw[black!50,thick] ($(AE.north west)+(-1,0.6)$)  rectangle  ($(AEff.south east)+(1,-0.6)$) node[anchor=south, pos=0.5, yshift=3cm] {\textbf{Performance}};
\draw[black!50,thick, fill=black!30] ($(PEOU.north west)+(-1,0.6)$)  rectangle ($(PU.south east)+(1,-0.6)$) node[anchor=south, pos=0.5, yshift=2.925cm] {\textbf{Perceptions}};
\draw[black!50,thick] ($(IU.north west)+(-1,2.1)$)  rectangle ($(IU.south east)+(1,-2.1)$) node[anchor=south, pos=0.5, yshift=3cm] {\textbf{Intentions}};
\draw[black!50,thick, fill=black!30] ($(AU.north west)+(-1,2.1)$)  rectangle ($(AU.south east)+(1,-2.1)$) node[anchor=south, pos=0.5, yshift=3cm] {\textbf{Behavior}};

    \draw [arrow] (AE) -- (PEOU);
    \draw [arrow] (AEff) -- (PU);
    \draw [arrow] (PEOU) -- (PU);
    \draw [arrow] (PEOU) -- (IU);
    \draw [arrow] (PU) -- (IU);
    \draw [arrow] (IU) -- (AU);

\node (AE) [memnode, xshift=-3cm] {\NodeTextSize\shortstack{Actual\\Efficiency}};
\node (AEff) [memnode, below of=AE] {\NodeTextSize\shortstack{Actual\\{\smallerNodeText Effectiveness}}};%\\\hypothesis{1}, \hypothesis{1}, \hypothesis{1}} };
\node (PEOU) [memnode, right of=AE, xshift=.75cm]  {\NodeTextSize\shortstack{Perceived\\Ease of\\Use}};%\\\hypothesis{1}, \hypothesis{1}} };
\node (PU) [memnode, below of=PEOU] {\NodeTextSize\shortstack{Perceived\\Usefulness}};%\\\hypothesis{1}, \hypothesis{1}} };
\node (IU) [memnode, right of=PEOU, yshift=-1.5cm, xshift=.75cm] {\NodeTextSize\shortstack{Intention\\to Use}};%\\\hypothesis{1} }};
\node (AU) [memnode, right of=IU, xshift=.75cm] {\NodeTextSize\shortstack{Actual\\Usage} };

\end{tikzpicture}}
    \caption{The Method Evaluation Model (MEM)~\cite{moodyMethodEvaluationModel2003}.}
    \label{fig:mem}
    \end{figure}

\subsection{Theoretical Framework}
\label{ssec:methods-theory}

Our study is built around the Method Evaluation Model (MEM), shown visually in Figure~\ref{fig:mem}, which describes the effect of actual and perceived performance on intention and actual use~\cite{moodyMethodEvaluationModel2003}. This is a common theoretical model~\cite{borstler2024acceptance} around which many acceptability studies have been designed~\cite{labunetsSecurityRiskAssessmentThesis, liinasuoChoosingRemoteFacetoface2023,  massacciAssessingRequirementsEvolution2014, zarraonandiaUsingCombinatorialCreativity2017, cardenas-delgadoMethodCreateMicrolearning2022, benbelkacemIntegratingHumanComputer2014}. The MEM consists of six components: Actual Efficiency (AEffic), Actual Effectiveness (AEffec), Perceived Usefulness (PU), Perceived Ease of Use (PEOU), Intention to Use (ITU), and Actual Use (AU).  %The MEM is a model that describes the effect of actual and perceived performance of a method on the intention and actual use of that method~\cite{moodyMethodEvaluationModel2003}. 

We assess the elements of the MEM in the following ways: \textbf{AEffic.} We measure AEffic by asking participants how long they spent on the task and how many versions of the model they created. We normalize these values according to the procedure described in Section~\ref{sssec:results-aa-aeffic}. \textbf{AEffec.} We measure AEffec by assessing the actual models submitted according to a pre-defined rubric. We qualitatively assess each model according to whether it follows the construction rules, whether the model is clearly conveying the attack information, and whether the model shows an analysis of the attack scenario. AEffic and AEffec are only measured for AA. \textbf{PU/PEOU/ITU.} We measure PU, PEOU, and ITU using a perception questionnaire that combines Likert and ranking questions. PU is measured with questions \texttt{L1}, \texttt{L2}, \texttt{L3}, \rank{1}, \rank{2}, and \rank{3}. PEOU is measured with questions \texttt{L4}, \texttt{L5}, \rank{4} and \rank{5}. ITU is measured with question \rank{6}. We address why ITU is only measured in terms of CPA in Section~\ref{sec:limitations}. The question text and evaluation rubric are provided in our supplementary material (see Section~\ref{sec:data}).

% \og{you need to provide more details about the MEM (why it is even useful for this study? What are the components that we measure? And also provide forward pointers explaining how each measured component will be operationalized (e.g., "We explain how we assess Actual Efficiency (AE) in Section V-B, etc.")}\NS{I'm going to do it here, and we can move it if it fits elsewhere.} 

\subsection{Study Design}
\label{ssec:methods-study-design}

Our study was implemented as an assignment in an introductory cybersecurity course. The assignment was a mandatory, graded assignment; students in the course had to explicitly consent for their assignment submission to be used in the study. We discuss the ethical considerations of the study design in Section~\ref{ssec:methods-ethics}. 

The assignment was designed to be reflective of how security practitioners apply threat modeling and risk assessment in practice. Participants were told to assess their threat scenarios as though they were a security analyst: they needed to understand how a threat happened and to be able to communicate this information to others, as this is broadly representative of how threat modeling can be used in practice~\cite{schneierSecretsLiesDigital2000,verreydt2024threat}. 

% Specifically, the assignment consisted of having students read provided news articles concerning a recent cyber attack. The student would then create a threat model using only the information found in the provided texts or that could be reasonably inferred from the text. We provided three examples of cyber attacks (each with three provided news articles), and the three selected threat models to examine in our study. 

All students were given the same three scenarios and were assigned a threat model for each scenario following a Latin squares design. We provided three news articles for each of the scenarios. Participants were instructed to use the information contained in the articles and any information that could reasonably be inferred from them. The models and scenarios are discussed in detail in Section~\ref{ssec:methods-models-scenarios} and the assignment of scenarios to participants is discussed in Section~\ref{ssec:methods-participants-training}.
% Students were assigned threat models and news articles based on a Latin squares design using the last digit of their student identification number. 
% The Latin squares assignment is provided in Table~\ref{tab:latin-squares-assignment}. 
% Figure~\ref{fig:student-number-last-digit-counts} shows the distribution of last digits of student numbers. Table~\ref{tab:threats-counts} shows the number of threats identified by each threat model for each scenario.

% Due to the constraints of the study being an assignment, specific controlled conditions were impossible to achieve. In our questionnaire, we ask students an objective question about actual efficiency, but we cannot precisely measure these with our study design. We can, however, accurately measure perceptions through perception questions, and we can still measures actual efficiency by asking students an objective questions (``how long did this take?'' and ``how many versions did you create?''). Using the MEM, we can use these measurements of perception to better understand the intention of our participants to use each of these threat models. We can use the submitted models and qualitatively analyze these models to understand the actual effectiveness of the models. Our qualitative rubric is provided in Appendix~??, and the qualitative evaluations were performed by two evaluators independently.

Participants were not required to use any specific tools to complete these tasks. While tools for creating ADTs, such as ADTool\footnote{\url{https://satoss.uni.lu/members/piotr/adtool/}} and ADTapp\footnote{\url{https://adtweb.app}}, and a tool for creating CORAS diagrams\footnote{\url{https://coras.tools}} exist, we could not find a convenient and usable tool for designing AGs. We provided links to the existing tools for participants, but made it clear to all participants that they were not required to use these tools. We ensured to clarify to all participants that a handwritten submission, provided it was legible, would be a valid submission.

\subsection{Models and Scenarios Selected}
\label{ssec:methods-models-scenarios}

As mentioned, the three models, ADTs, AGs, and CORAS, were selected as they are general graphical threat models. They are \emph{general} as they can be applied to any scenario, even a non-technical or physical threat~\cite{mauwFoundationsAttackTrees2006,kordyFoundationsAttackDefense2011,sheynerAutomatedGenerationAnalysis2002,solhaugCORASLanguageWhy2014}. This is in contrast to threat models such as STRIDE, which are specifically designed to elicit threats against computer systems~\cite{kohnfelder1999threats}. They are \emph{graphical} in that they are made up of nodes and edges, and they present information visually. While other graphical threat models exist, we selected these three as they all organize information differently: component-based, state-based, and aspect-based, respectively. We believe that these models are representative of the broad array of possible graphical threat models~\cite{haqueEvolutionaryApproachAttack}.

%While the three selected scenarios cannot be a representative sample of all possible cyber attacks, we selected scenarios that were considerably different. Our intention was to be able to find if a difference existed in the actual and perceived usage of the threat models between the scenarios, not to be able to precisely clarify this difference. \og{the previous sentence is not clear. Better to just state again what you intend to find in the context of the scenarios} 

We selected three diverse attack scenarios: a financial attack on a bank (\emph{Revolut}),
% \footnote{\url{https://thefintechtimes.com/revolut-card-fraud-dropped-by-30-since-scam-detection-feature-launch/}}
% \footnote{\url{https://www.telegraph.co.uk/money/banking/revolut-fraud-payout-row-leaves-1000-victims-in-limbo/}}
% \footnote{\url{https://www.theguardian.com/money/2024/apr/10/im-a-victim-of-scammers-but-revolut-says-no-to-a-refund}}
a personal data leak at a genetic testing company 
(\emph{23 and me}),
% \footnote{\url{https://www.bbc.com/future/article/20240212-dna-testing-what-happens-if-your-genetic-data-is-hacked}}
% \footnote{\url{https://www.nbcnews.com/tech/security/23andme-hackers-gained-access-data-69-million-people-rcna127990}}
% \footnote{\url{https://www.nytimes.com/2023/12/04/us/23andme-hack-data.html}}
and an account data breach at a cloud storage company (\emph{Dropbox}).
% \footnote{\url{https://www.securityweek.com/dropbox-data-breach-impacts-customer-information/}}
% \footnote{\url{https://www.itpro.com/security/the-dropbox-data-breach-is-a-classic-case-of-breach-by-acquisition}}
% \footnote{\url{https://www.forbes.com/sites/daveywinder/2024/05/02/dropbox-warns-hacker-accessed-customer-passwords-and-mfa-data/}}
 These scenarios were selected as they all had different types of attacks, and all had different types of data that needed to be protected. If a difference according to the MEM is found, that would be evidence that some threat models could be better suited for some scenarios, and more research would be needed to define which models work for which scenarios and why. No difference would be evidence that general graphical threat models can be used regardless of the scenario.

\subsection{Data Processing and Analysis}
\label{ssec:methods-data-processing}

As we have at least three groups (one for each threat model) for each question, multiple-group statistical testing is required for our analysis. Kruskal-Wallis (KW) test has been used previously by Labunets et al.~\cite{labunets2017model} in a study with a similar design to ours that compared tabular and graphical threat models, and in other similar studies, e.g., Naiakshina~\etal~\cite{naiakshinaConductingSecurityDeveloper2020} and Katta~\etal~\cite{kattaComparingTwoTechniques2010}. KW is a statistical test for checking if one group, in a set of three or more groups, stochastically dominates the other groups~\cite{dunnMultipleComparisonsUsing1964}. 
KW is a non-parametric test, which suits our data as the Shapiro-Wilk test shows our data is not normally distributed~\cite{hanuszShapiroWilkTest2016}. Further, using the Kolmogorov–Smirnov test, we find that our groups have sufficiently similar distributions~\cite{simardComputingTwoSidedKolmogorovSmirnov2011}. The KW test requires that samples are independent, and we assume this independence in our data.

The KW test will only tell us if there is a significant difference between the groups, not which groups are different. To determine which groups are different, we apply the Conover test~\cite{conoverMultiplecomparisonsProceduresInformal1979}. The Conover test is a post-hoc test for the KW test, which will tell us which groups are different. The Conover test is a common post-hoc test for the KW test, and has been used in similar studies~\cite{labunetsExperimentalComparisonTwo2013, gonzalezAutomatedCharacterizationSoftware2019, scandariatoEmpiricalAssessmentSecurity2014}.

If the KW test does not find a significant difference between the groups, we apply the Wellek-Welch (WW) test to determine if there is significant equivalence between the groups. The Wellek-Welch test is an extension of the non-parametric Welch test, which allows for equivalence testing in three or more groups~\cite{kohRobustTestsEquivalence2013}. We only apply the Wellek-Welch test if the KW test does not find a significant difference between the groups. This structure of testing has been used previously in similar studies, albeit those conducting difference/equivalence testing in only two groups~\cite{labunets2017model, labunets_no_2018, broccia_assessing_2024, broccia2025evaluating,schieleAcceptability2025}. 

We use a significance threshold, $\alpha$, of 0.05 for all tests. To account for family-wise error rate (FWER), we apply the Holm-Bonferroni (HB) correction to our $p$-values. The HB correction divides $\alpha$ by $m$, where $m$ is the number of tests with a smaller $p$-value. For simplicity of representing our results, we report $p$-values that have been multiplied by $m$. This moves $m$ to the other side of the inequality and allows us to present modified $p$-values that have the significance threshold of 0.05.

\subsection{Participants and Training}
\label{ssec:methods-participants-training}

We recruited 38 student participants for the study out of the total course population of 55 students. Participants were third-year bachelor students studying varying subjects, e.g., related to international relations, political science, and public administration, at Leiden University (NL). The students participated in a minor related to cybersecurity governance. At the time of the study, they had received a limited (approx. 2 months) education on cybersecurity and developed basic programming skills. Therefore, similarly to~\cite{schieleAcceptability2025}, we can conclude that our participants had a limited technical background. 

The training on threat models took place in October 2024 during a 2-hour lecture on threat models. All three threat models were introduced, and students performed an in-class exercise to create their own version of each type of threat model. Attendance was not required, but highly encouraged. The lecture was given by the first author of this paper. The training was designed by the authors of this paper who are experienced educators in threat modeling and cybersecurity in general.

\begin{figure}
    \includegraphics[width=\linewidth]{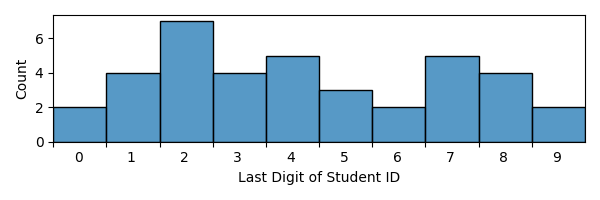}
    \caption{Histogram of last digits of participant student numbers.}
    \label{fig:student-number-last-digit-counts}
\end{figure}

In Table~\ref{tab:threats-counts}, we see the distribution of threat models and the scenarios assigned. The scenarios are not evenly distributed. This is an artifact from the Latin squares design, where two numbers were assigned the same Latin squares assignment (there are ten possible digits, but only nine scenarios), and the number two is disproportionately represented in the participant population. This is a limitation of the study design, and we will have to account for these results in our analysis. This is discussed further in Section~\ref{sec:limitations}.

Our participants self-selected to participate in the study. Evaluating the final grades of the participants against non-participants, we find that participants had a final average grade of 66.4\%, while non-participants had a final average grade of 44.9\%. The Kruskal-Wallis test shows a statistical difference between the two groups ($p=0.004$). This indicates that participants were more engaged with the course than non-participants, and the self-selection might have introduced bias. We discuss this further in Section~\ref{sec:limitations}.

\begin{table}
    \centering
    \caption{Comparison of the number of threats identified by each threat model for each scenario.}
    \label{tab:threats-counts}
    \begin{tabular}{lrrr}
        \toprule
        &\multicolumn{3}{c}{\textbf{Scenarios Assigned}}\\
        \textbf{Threat Model} &  Revolut &  23 and me &  Dropbox \\
        \midrule
        ADT          &       10 &         17 &       11 \\
        Attack Graph &       11 &         10 &       17 \\
        CORAS        &       17 &         11 &       10 \\
        \bottomrule
        \end{tabular}
\end{table}

\subsection{Ethical Considerations}
\label{ssec:methods-ethics}

Collecting data from a mandatory graded assignment invokes significant and important ethical considerations: namely, how to ensure that students do not feel coerced to participate. We have taken numerous safeguards to ensure students were not placed under any pressure to participate. Our study design was reviewed and approved by the relevant Research Ethics Committee at Leiden University. 

Our safeguards included having trained teaching assistants (TAs) grade the assignment according to a pre-designed rubric. The TAs did not know who were participants or non-participants, and there was no entry on the rubric that could be influenced by participation. We provided a way to submit the signed consent form via an independent third party, so that we (the study authors) would not know who submitted the consent form. Further, we provided an online consent withdrawal form which could be filled it at any time with minimal effort and time. We informed the students that study responses would not be collected until one month after the final grade submission deadline, after which it is not possible to change grades. We explicitly provided the procedure of initially submitting consent and withdrawing it after receiving the final grade to any student who was concerned about study participation affecting their grade. Further, we provided contact information for independent personnel within the university that students could refer to with any complaint or if they were seeking advice. No student followed the consent withdrawal procedure or contacted the third-party oversight personnel.

We followed the guidelines in the Menlo Report~\cite{kenneallyMenloReportEthical2012}, specifically focusing on balancing of benefits against potential harms. Overall, as our students were interested in cybersecurity, practicing creating threat models is a valuable benefit to them. Further, the study offers valuable insight into the perception of threat models that further develops empirical insights into threat modeling. As such, we believe the study creates more benefits for the participants and the community than potential harms. Additionally, as a small compensation for participants' time, we distributed three 20\euro{} gift cards in a raffle.

\section{Results}
\label{sec:results}

\subsection{\hypothesis{1}: Absolute Acceptability (AA) of Threat Models}
\label{ssec:results-likert_overall}

In Table~\ref{tab:overall_likert_tests}, we present the results of statistical tests run on our qualitative evaluations of the threat models (AEffec), the reported time taken and different versions created by participants (AEffic), and the perception of usefulness (PU) and ease of use (PEOU) from the Likert items. We cover all these elements in Section~\ref{ssec:methods-theory}.

\subsubsection{Actual efficiency (AEffic)}
\label{sssec:results-aa-aeffic}
As discussed in \ref{ssec:methods-theory}, we could not directly collect the time taken by participants. We asked open-ended questions on how long it took them to create the threat models and how many versions they created. Participants differed in their interpretation of this question, with some answering how long it took only to draw each threat model, and others answering how long it took to read the scenarios, analyze the attack scenario, and then create the threat model. This was also the case for versions, where some participants interpreted significant modification as ``creating a new version'' while others only considered a new version to be a completely new model. We normalized these data by presenting values as a percentage of the largest value provided by each participant. In this way, the interpretation of the student did not matter, as the normalized data could be subsequently compared. This method of processing the data will be valid provided each student answered the question the same way for each threat model, which we believe is a reasonable assumption. We found that for both actual efficiency questions, there was significant equivalence between the threat models. This suggests that the actual effort needed to create the threat models was not affected by the threat model selected.

\subsubsection{Actual effectiveness (AEffec)}
\label{sssec:results-aa-aeffec}
Participants differed significantly in following the construction rules for AGs (specifically, AGs differ significantly compared to CORAS). As this difference is only between AGs and CORAS, and we can see from our evaluations that AGs were generally constructed more poorly, while CORAS was the most correctly constructed model. Several of the poor evaluations of AGs came from participants who did not appear to fully understand the difference between AGs and ADTs, and thus poorly constructed both models. This issue did not seem to occur with CORAS. In our qualitative evaluations of the submitted models as a means of analysis and communication, we evaluated the models to be significantly equivalent across all scenarios. This suggests the actual effectiveness of the models is the same for all participants.

\subsubsection{Perceived usefulness (PU)}
\begin{figure}
    \centering
    \includegraphics[width=0.45\textwidth]{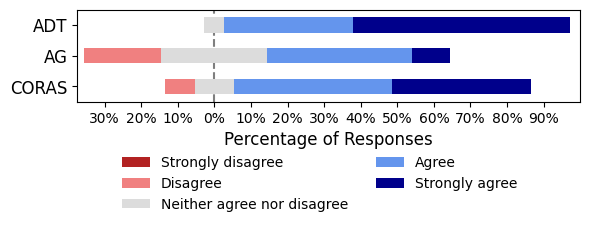}
    \caption{Distribution of responses to the Likert statement ``I find this model to be an effective threat model'' (\texttt{L1}) per threat model given as a percentage of the total number of responses for each model.}
    \label{fig:L1}
\end{figure}

For the Likert items related to perceived usefulness, we found there was a significant difference w.r.t. general usefulness of the model (\texttt{L1}). The Conover post-hoc test shows there is a significant difference between AGs and the other two models. Looking at the data shown in Figure~\ref{fig:L1}, we see the AGs were evaluated significantly lower than the other two models. However, when asking participants about the two primary uses of threat models (analysis and communication, \texttt{L2} and \texttt{L3}, respectively), we found significant equivalence between the threat models. A possible explanation for this result is that AGs are the only model without an associated tool. Participants were not required to use a tool to create any of the threat models; however, online tools for ADTs and CORAS exist, and participants were told about the presence of these tools.

\subsubsection{Perceived ease of use (PEOU)}
For the Likert items related to perceived ease of use, we found that there was significant equivalence for PEOU for both questions. This suggests that the perceived ease of use of the threat models was not affected by the threat model selected. This follows from the results of actual efficiency. Our participants did not experience or perceive a difference in the effort used to create the threat models.

    \begin{table*}[t]
        \centering       
        \caption{Statistical analysis of Likert scale questions and time normalized. %\texttt{L1}, \ texttt {L2}, and \texttt{L3} are questions related to the PU, and \texttt{L4}  \texttt{L5} are related to the PEOU.
        }
        \label{tab:overall_likert_tests}
    
\resizebox{.8\linewidth}{!}{
\begin{tabular}{lllllcccl}
            \toprule
          \textbf{MEM} & \textbf{Question} & \textbf{KW Test} & \textbf{WW Test} & \multicolumn{3}{c}{\textbf{Conover Test}}  \\
           &  &  &  & ADTs -- AGs & AGs -- CORAS & CORAS -- ADTs  \\\midrule
          \multirow{3}{*}{{\shortstack[l]{Actual\\Effectiveness}}} & Construction & \textbf{8.01e-03} &  & 0.125 & \textbf{1.41e-03} & 1.0  \\
           & Analysis & 1.0 & \textbf{9.27e-07} &  &  &   \\
           & Communication & 1.0 & \textbf{7.25e-04} &  &  &   \\\hdashline
          \multirow{2}{*}{\shortstack[l]{Actual\\Efficiency}} & Time & 1.0 & \textbf{1.49e-03} &  &  &   \\
           & Versions & 1.0 & \textbf{5.84e-04} &  &  &   \\\hdashline
          \multirow{3}{*}{{PU}} & \texttt{L1} & \textbf{3.52e-05} &  & \textbf{1.17e-06} & \textbf{9.91e-03} & 0.366  \\
           & \texttt{L2} & 0.47 & \textbf{7.31e-03} &  &  &   \\
           & \texttt{L3} & 0.19 & \textbf{3.68e-02} &  &  &   \\\hdashline
          \multirow{2}{*}{{PEOU}} & L4 & 1.0 & \textbf{3.92e-06} &  &  &   \\
           & \texttt{L5} & 1.0 & \textbf{1.32e-07} &  &  &   \\      
        \bottomrule
        \end{tabular}
}
    \end{table*}

\highlight{\textbf{\hypothesis{1}}: Eight out of ten tests show significant equivalence between the threat models. Of the two showing a significant difference, one is in AEffec and one in PU, both being one test out of three for their respective component of the MEM.}

\subsection{\hypothesis{2}: Comparative Perception of Acceptability (CPA) between Threat Models}
\label{ssec:results-ranking_overall}

    \begin{table*}[t]
        \centering       
        \caption{Statistical analysis of ranking questions.}% \rank{1}, \rank{2}, and \rank{3} are related to the perceived usefulness, \rank{4} and \rank{5} are related to the perceived ease of use, and \rank{6} is related to the intention to use.}
        \label{tab:overall_rank_tests}
    
\resizebox{.76\linewidth}{!}{
\begin{tabular}{lcllcccl}
            \toprule
          \textbf{MEM} & \textbf{Question} & \textbf{KW Test} & \textbf{WW Test} & \multicolumn{3}{c}{\textbf{Conover Test}}  \\
           &  &  &  & ADTs -- AGs & AGs -- CORAS & CORAS -- ADTs  \\
           \midrule
          \multirow{3}{*}{{\shortstack[l]{PU}}} & \rank{1} & \textbf{1.66e-07} &  & \textbf{1.73e-09} & \textbf{1.16e-05} & 0.807  \\
           & \rank{2} & \textbf{5.25e-07} &  & \textbf{3.02e-09} & \textbf{9.22e-04} & 0.109  \\
           & \rank{3} & \textbf{2.91e-03} &  & \textbf{4.52e-02} & \textbf{5.26e-04} & 1.0  \\\hdashline
          \multirow{2}{*}{{PEOU}} & \rank{4} & 1.0 & \textbf{0.00e+00} &  &  &   \\
           & \rank{5} & 0.392 & \textbf{4.23e-04} &  &  &   \\\hdashline
          \multirow{1}{*}{{ITU}} & \rank{6} & 0.063 & 0.158 &  &  &   \\      
        \bottomrule
        \end{tabular}
}
    \end{table*}

In Table~\ref{tab:overall_rank_tests}, we present the results of the statistical tests on the ranking data. Participants were asked to rank each of the threat models according to the different MEM dimensions.

\begin{figure*}[ht]
    \centering
    \begin{subfigure}[b]{0.32\textwidth}
        \centering
        \includegraphics[width=\textwidth]{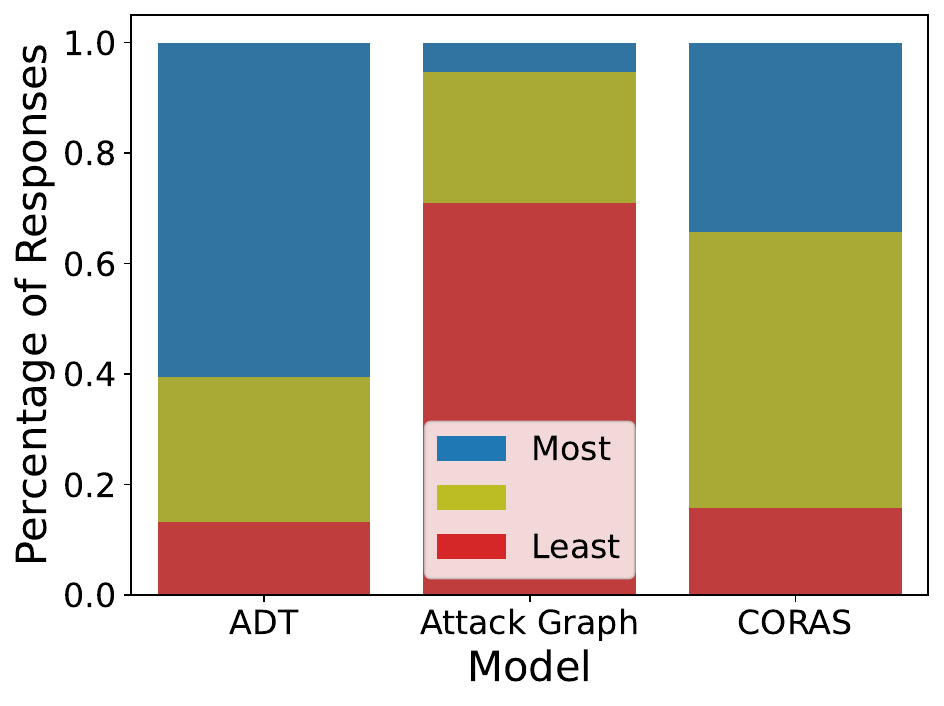}
        \caption{Rankings of the perceived general effectiveness of the threat models (\rank{1}) \text{  }\text{   }\text{  }\text{   }\text{  }\text{   }\text{  }\text{   }\text{  }\text{   }\text{  }\text{   }\text{  }\text{   }\text{  }\text{   }\text{  }\text{   }\text{  }\text{   }}
        \label{fig:rank1}
    \end{subfigure}
    \hfill
    \begin{subfigure}[b]{0.32\textwidth}
        \centering
        \includegraphics[width=\textwidth]{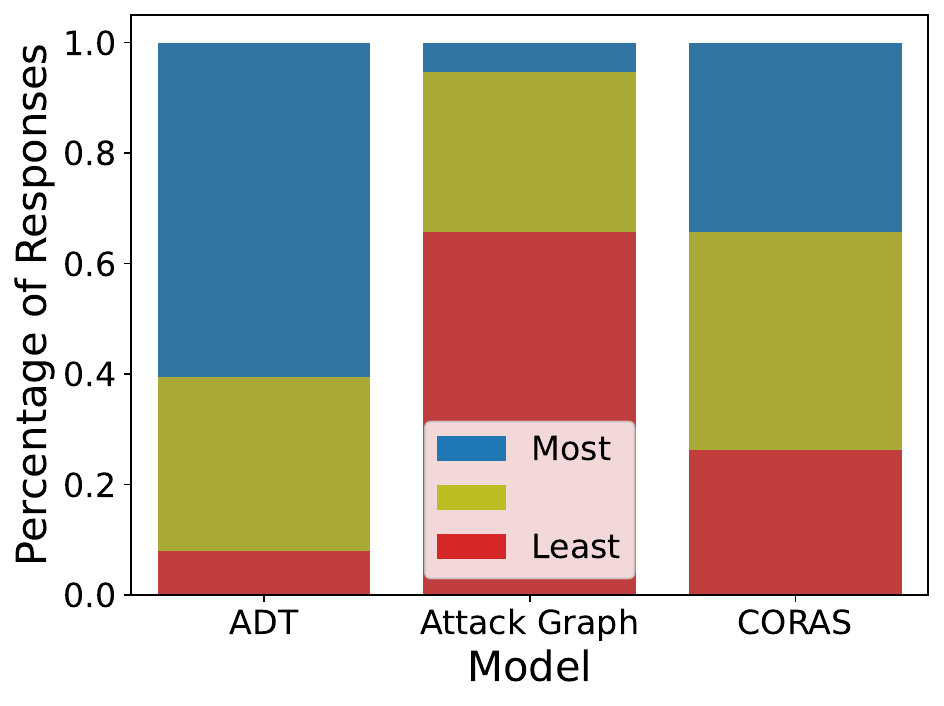}
        \caption{Rankings of the perceived effectiveness of the threat models as a means of analysis (\rank{2})}
        \label{fig:rank2}
    \end{subfigure}
    \hfill
    \begin{subfigure}[b]{0.32\textwidth}
        \centering
        \includegraphics[width=\textwidth]{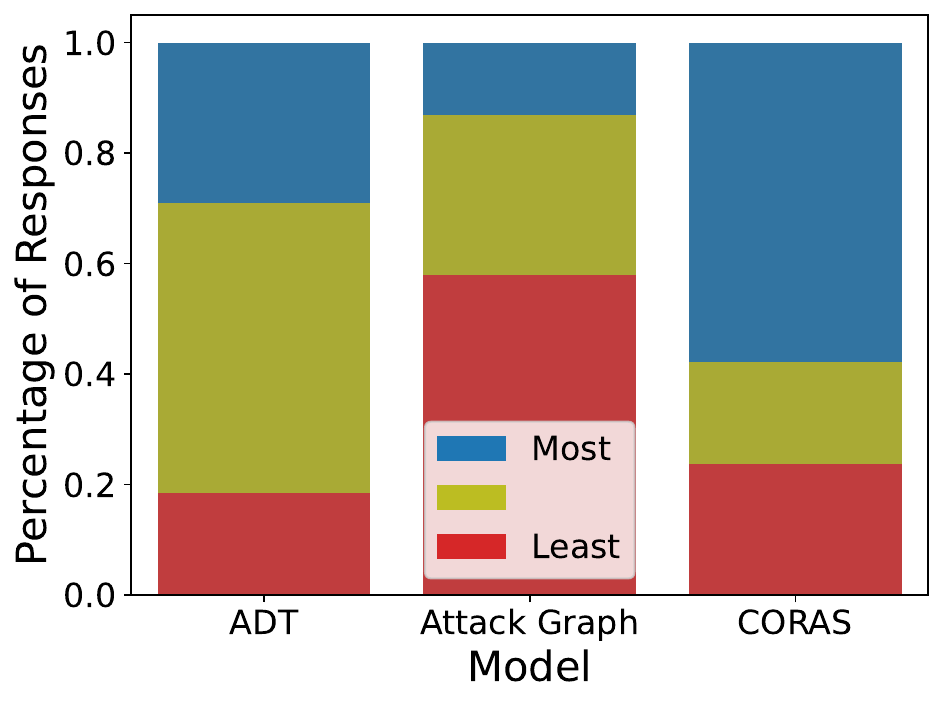}
        \caption{Rankings of the perceived effectiveness of the threat models as a means of communication (\rank{3})}
        \label{fig:rank3}
    \end{subfigure}
    \caption{Rankings of the perceived effectiveness of the threat models}
    \label{fig:ranks123}
\end{figure*}

Of note in our results are rankings \texttt{R1}, \texttt{R2}, and \texttt{R3}, which are all related to PU. The KW test shows that at least one of the three models stochastically dominates the others, but does not indicate which. Because of this result, we apply the Conover post-hoc test in accordance with our data analysis plan. We find that AGs differ significantly from both ADTs and CORAS in all three PU rankings. When we examine the percentage distribution of the three rankings in Figure~\ref{fig:ranks123}, we see that AGs are consistently ranked lower than the other two models. From the previous section, we saw that AGs were absolutely perceived as generally less acceptable than the other two models, and we have strong evidence that it is perceived as less acceptable compared to ADTs and CORAS.

The results show that for PEOU, participants did not consistently rank threat models differently. In fact, for \rank{5} (how long did it take participants), the distribution of responses across our sample is identical: all threat models received the same total amount of each ranking. This shows that for PEOU, there is no comparative difference for our participants; otherwise put, we find no evidence that one model is significantly easier to use. For ITU, we found neither a significant difference nor a significant equivalence.

\highlight{\textbf{\hypothesis{2}}: The comparative PEOU of ADTs, AGs, and CORAS is significantly equivalent. The comparative PU shows a difference for AGs, which were perceived as comparatively less useful than ADTs and CORAS. No conclusion can be made for ITU.}

\subsection{\hypothesis{3}: Absolute Acceptability between Scenarios}
\label{ssec:results-likert_by_model_within_scenario}

    \begin{table}[t]
        \centering       
        \caption{Statistical analysis of Likert scale questions and time normalized. L1, L2, and L3 are questions related to the perceived usefulness and L4, L5, and normalized time are related to the perceived ease of use.}
        \label{tab:likert_statistical_tests_models}
    
\resizebox{\linewidth}{!}{
\begin{tabular}{llllll}
            \toprule
          \textbf{MEM} & \textbf{Question} & \textbf{Model} & \textbf{KW Test} & \textbf{WW Test}  \\\midrule
          \multirow{9}{*}{{\shortstack[l]{Actual\\Effectiveness}}} & \multirow{3}{*}{\shortstack[l]{Construction}} & ADT & 1.0 & 0.39  \\
           &  & AG & 1.0 & 0.808  \\
           &  & CORAS & 1.0 & \textbf{4.80e-02}  \\\cdashline{2-5}
           & \multirow{3}{*}{\shortstack[l]{Analysis}} & ADT & 1.0 & \textbf{2.72e-02}  \\
           &  & AG & 1.0 & 0.083  \\
           &  & CORAS & 1.0 & 0.184  \\\cdashline{2-5}
           & \multirow{3}{*}{\shortstack[l]{Communication}} & ADT & 1.0 & \textbf{4.93e-02}  \\
           &  & AG & 1.0 & \textbf{1.39e-02}  \\
           &  & CORAS & 1.0 & \textbf{0.00e+00}  \\\cdashline{1-5}
          \multirow{6}{*}{\shortstack[l]{Actual\\Efficacy}} & \multirow{3}{*}{\shortstack[l]{Time}} & ADT & 0.828 & 1.0  \\
           &  & AG & 1.0 & \textbf{4.48e-02}  \\
           &  & CORAS & 1.0 & 1.0  \\\cdashline{2-5}
           & \multirow{3}{*}{\shortstack[l]{Versions}} & ADT & 1.0 & 0.201  \\
           &  & AG & 1.0 & 0.224  \\
           &  & CORAS & 1.0 & 0.063  \\\cdashline{1-5}
          \multirow{9}{*}{{PU}} & \multirow{3}{*}{L1} & ADT & 1.0 & 0.823  \\
           &  & AG & 1.0 & 0.562  \\
           &  & CORAS & 0.312 & 1.0  \\\cdashline{2-5}
           & \multirow{3}{*}{L2} & ADT & 1.0 & 0.145  \\
           &  & AG & 1.0 & \textbf{4.24e-03}  \\
           &  & CORAS & 1.0 & \textbf{1.22e-03}  \\\cdashline{2-5}
           & \multirow{3}{*}{L3} & ADT & 1.0 & 1.0  \\
           &  & AG & 1.0 & 0.095  \\
           &  & CORAS & 1.0 & 1.0  \\\cdashline{1-5}
          \multirow{6}{*}{{PEOU}} & \multirow{3}{*}{L4} & ADT & 1.0 & 0.289  \\
           &  & AG & 1.0 & \textbf{4.60e-02}  \\
           &  & CORAS & 1.0 & \textbf{4.67e-02}  \\\cdashline{2-5}
           & \multirow{3}{*}{L5} & ADT & 1.0 & 0.632  \\
           &  & AG & 1.0 & 0.323  \\
           &  & CORAS & 1.0 & \textbf{4.23e-03}  \\      
        \bottomrule
        \end{tabular}
}
    \end{table}
\begin{figure}[ht]
    \centering
    \begin{subfigure}[b]{\linewidth}
        \centering
        \includegraphics[width=\linewidth]{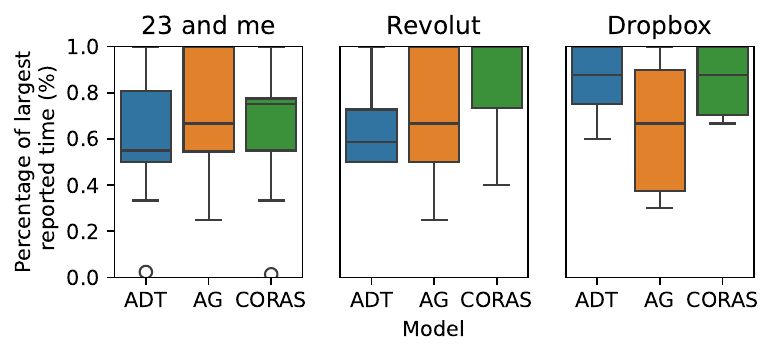}
        \caption{Normalized time taken for each scenario per threat model}
        \label{fig:time_per_scenario}
    \end{subfigure}
    \hfill
    \begin{subfigure}[b]{\linewidth}
        \centering
        \includegraphics[width=\linewidth]{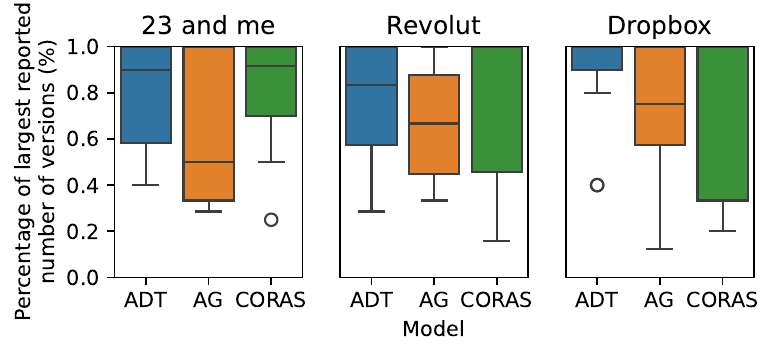}
        \caption{Normalized number of versions created for each scenario per threat model}
        \label{fig:version_per_scenario}
    \end{subfigure}
    \caption{Actual efficiency metrics per scenario per model}
    \label{fig:actual_efficiency_per_scenario}
\end{figure}

% \begin{figure}
%     \centering
%     \includegraphics[width=0.45\textwidth]{Code/img/Time_by_scenario.png}
%     \caption{Normalized time taken for each scenario per threat model}
%     \label{fig:time_per_scenario}
% \end{figure}

In Table~\ref{tab:likert_statistical_tests_models} we present the same results as in Table~\ref{tab:overall_likert_tests}, but now broken down by scenario. These comparisons are within each model, grouping participants based on which scenario they were assigned for each model. Not a single significant difference was found in our evaluations, the normalized time taken, or in any of the Likert items. This suggests that the scenarios did not have an effect on the acceptability of the threat models.

For AEffec, we see a significant equivalence for CORAS w.r.t. following construction rules, and ADTs w.r.t. using the model for analysis. For using the models as a means of communication, all models were found to be significantly equivalent (with CORAS results being identical). This is significant evidence that the actual effectiveness of the models is not affected by the scenarios. For AEffic, we see that there is significant equivalence found for AG w.r.t. time taken, suggesting that AG can be consistently applied at a similar speed across different scenarios. We do not see significant results for ADTs and CORAS, nor do we find any significant results when examining the number of versions of each model created. precluding us from concluding whether the time taken to create these models is affected by the scenarios. We show this data in Figure~\ref{fig:actual_efficiency_per_scenario}, and it is clear that no consistent pattern emerges.

Regarding PU, we do not see significant results for the general usefulness of the models (\texttt{L1}) and the usefulness of the models as a means of communication (\texttt{L3}). However, we do see significant equivalence for both CORAS and AGs w.r.t. the perceived usefulness of the models as a means of analysis (\texttt{L2}). This suggests that the scenario selected does not affect the perception of the models as a means of analysis.

Finally, regarding PEOU, we see significant equivalence for CORAS on both Likert items, suggesting that CORAS was consistently perceived by our participants as easy to use regardless of the scenario. AG was perceived as easy to use regardless of scenario in \texttt{L4}, but the same perception did not hold when it came to whether the model enabled the participant to complete the task more quickly (\texttt{L5}). ADTs did not have significant results for either question.

\highlight{\textbf{\hypothesis{3}}: There is no evidence to suggest that the acceptability of ADTs, AGs, and CORAS is affected by the scenarios.}

\subsection{Difference in Tool Use and Perception}
\label{ssec:results-tool-use}

As seen previously, when significant differences exist, those differences are always between AGs and ADTs/CORAS. Within our study, AGs are a model without an associated tool, while ADTs and CORAS have online model-specific tools available. We asked participants in question \texttt{W4}, ``Did the availability/non-availability of a creation tool affect your ability to create this threat model?''. This question gives insight into whether the tools available made a difference in the acceptability of the threat models.

As this was an open-ended question, we coded the responses. We assigned a Likert code ranging from 1 to 3 (with 1 being disagree and 3 being agree) based on whether participants agreed that the availability/non-availability affected their ability to create the model. This code only reflected if there was a perceived effect, not if the effect was positive/negative. We see a significant difference between the three groups with a KW test $p$-value of \textbf{1.055e-05}. The Conover post-hoc test shows that AGs differ significantly from both ADTs and CORAS, with $p$-values of \textbf{4.195e-04} and \textbf{8.353e-07}, respectively. This difference is due to participants all agreeing that the purpose-built tools for CORAS and ADTs made the creation of these models easier, while for AG, participants did not believe the lack of a purpose-built tool affected their ability to create the models.

For the participants who answered that tool availability affected them, we assigned codes of 1 or -1, depending on whether the effect was positive or negative. The statistical analysis of these codes yielded the KW test $p$-value of \textbf{1.051e-11}. The Conover post-hoc test shows that AGs differ significantly from both ADTs and CORAS, with $p$-values of \textbf{2.397e-21} and \textbf{1.051e-22}, respectively. This is not surprising, as participants overwhelmingly agreed that the purpose-built tools aided their creation of the models, and no participant believed the lack of a tool was a boon to their ability to create a model. 

\begin{figure}[t]
    \centering
    \includegraphics[width=0.45\textwidth]{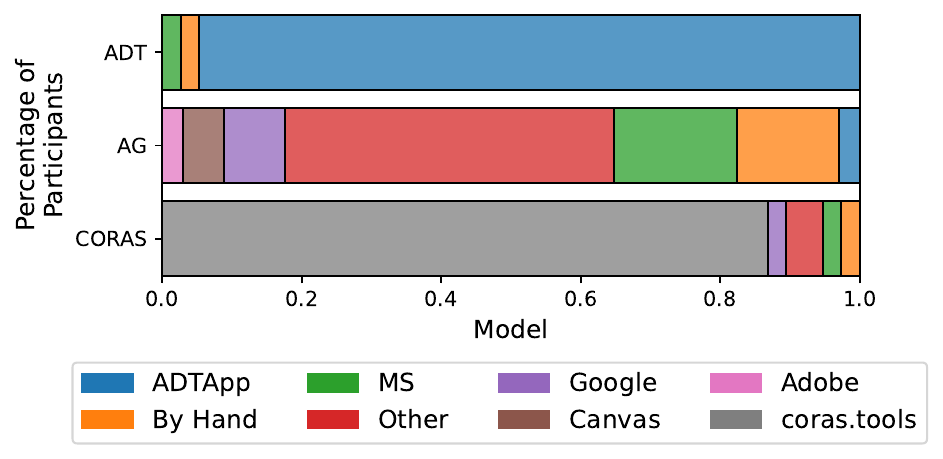}
    \caption{Distribution of tool use for the three models, according to responses and examining the produced output. MS tools included Visio, PowerPoint, and Word. Google tools included Google Drawings and Google Slides. ``Other'' was assigned when it was not clear which tool was used.}
    \label{fig:ToolTypes}
\end{figure}

Finally, we assess which tools participants used. Some participants explicitly stated which tool they used, and for many we could ascertain which tool was used by examining the model submission. For all other cases, we recorded the tool use as ``Other''. The distribution of tools used is provided in Figure~\ref{fig:ToolTypes}. We see that for ADTs and CORAS, the overwhelming majority of participants used the purpose-built tools. For AGs, there were a wide array of tools used.

\section{Discussion}
\label{sec:discussion}

\subsection{Acceptability of the Models}
\label{sec:discussion-acceptability}

Overall, the three graphical threat models were acceptable to our participants; there was significant equivalence for almost all measurements of aspects of the MEM. If we combine our results with those of Labunets~\etal~\cite{labunetsExperimentComparingTextual2014}, we find that while the acceptability of different types of models may differ depending on the situation, within at least one specific type of model (general, graphical threat models), the same distinction does not appear to hold. This is a positive result for the use of graphical threat models in industry, as it suggests that these threat models are acceptable regardless of how the models present information (component-based, state-based, or aspect-based for ADTs, AGs, and CORAS, respectively). For CPA, we found that AGs were perceived significantly worse than ADTs and CORAS. This is an interesting finding, as it suggests that AGs may be less acceptable than the other two models for the participants with a limited technical background. This is in contrast with Lallie et al.~\cite{lallieEmpiricalEvaluationEffectiveness2017} who found that for participants without a computer science background, AGs appear to be slightly more effective, compared to fault trees (in a non-significant way). However, our result may also be due to the lack of model-specific tooling for AGs.

\subsection{Effect of Scenarios}
\label{sec:discussion-scenarios}

A specific focus of our study was to understand if the scenario the model was used on would affect the acceptability of the model. Note that we were not attempting to specifically define which types of scenarios models would be acceptable for as this would require a different study design. 

Instead, we were interested in whether the scenario had any statistically significant effect on acceptability; this would suggest that such research into specific applicable scenarios would be necessary. We did not find any statistically significant differences in the acceptability of the models between scenarios. We found some statistically significant equivalences, but these were not consistent across all models. One important finding is that CORAS had the most significant equivalence between scenarios of any of the three models. This suggests that CORAS is the most robust w.r.t. which model is being used (that is, CORAS is best suited to model a wide array of different scenarios). A major reason why we did not see many statistically significant results is due to the relatively small sample size, the large number of tests (splitting the data by scenario and model exponentially increases the number of tests), and the conservative statistical correction (HB) applied. Our expectation would be that with a larger sample size, we would see more statistically significant equivalent results, but further research would be needed to confirm this.

\subsection{Effect of Tooling}
\label{sec:discussion-tooling}

While the use of model-specific tooling was not a primary focus of our study, the lack of a specific tool for AGs and the behavior of participants to predominantly use model-specific tooling when available (which was not a requirement; this was emergent behavior) did provide some insight into the potential effect of tooling. It is an interesting finding that when given the option, but not the requirement, to use a purpose-built tool, an overwhelming majority of participants elected to use that tool. This suggests that if a tool is available (and known) to users, they may gravitate toward using the specific tool. More research would be needed to understand if a propensity toward specific tooling is a general trend or if this is specific to our participants. As the tools provided to the participants did not have an associated threat library, we can preclude the possibility that the tools affected what was modeled.

All of the statistically equivalent differences we found were with AGs. AGs differed from ADTs and CORAS in our study, as AGs were the only model without a specific tool. This could suggest that the lack of a tool for AGs may have impacted the PU of AGs. It would be expected that tooling would affect the PEOU, as this speaks to the efficiency with which a user can create a model (how much effort is required); in general, our understanding is that the tool would aid users in creating the models faster or with less effort, thus improving both AEffic and PEOU. This sentiment is echoed by our participants in question \texttt{W5}, many of whom mentioned that the availability of model-specific tooling allowed them to create the ADT and CORAS models faster. However, the lack of such a tool for AGs did not significantly impact either AEffic or PEOU. More interesting still, AGs performed statistically significantly worse than ADTs and CORAS in PU. As we examined in Section~\ref{ssec:results-tool-use}, this may be due to the lack of specific tooling or due to AGs being perceived as less useful in general. More research would be needed to examine this further. It may be the case that tools have an outsized impact on PU instead of PEOU, which is not what we would have expected.

\subsection{Answer to the Research Question}

In general, the acceptability of the different general graphical threat models does not differ significantly. We see significant evidence that ADTs and CORAS are statistically equivalent w.r.t. the MEM. There are some differences with AGs specifically, but this may be due to the lack of model-specific tooling available in our study or due to some quality of AGs that makes them less useful. As our study was designed to be reflective of how threat models may be used in practice, our results strongly suggest that which threat model is specifically used is not as important as the fact that a threat model is used. This is an empirical result that supports a common finding in qualitative research on threat modeling~\cite{verreydt2024threat,kaurThreatModelingVery2025}.
\section{Limitations}
\label{sec:limitations}

Our study was conducted as a laboratory experiment, with the inherent limitations of this type of research~\cite{stol2018abc,storey2020software} being applicable. First, our study sample has several limitations. It was a sample of convenience, taken from the student population of one of our classes. Similar sampling has been performed in studies such as~\cite{labunetsExperimentComparingTextual2014,labunetsExperimentalComparisonTwo2013,opdahlExperimentalComparisonAttack2009,schieleAcceptability2025}, and has been shown to yield similar results compared to other types of participant samples~\cite{mullinixGeneralizabilitySurveyExperiments2015}. Druckman and Kam~\cite{druckmanStudentsExperimentalParticipants2011} empirically showed that samples of convenience only present a problem when the sample is specifically uniform on a studied metric, which is not the case for our study. Another point of concern is that students may not be representative of the target population (in this instance, threat modeling practitioners); however, some previous studies~\cite{karpatiInvestigatingSecurityThreats2015,naiakshinaConductingSecurityDeveloper2020} have shown that for computer science and cybersecurity, students are a reasonable proxy for practitioners. It may be the case that our study does not capture the organizational context in which these models may be used.

As mentioned in Section~\ref{ssec:methods-participants-training} and shown in Figure~\ref{fig:student-number-last-digit-counts}, our sample was not evenly distributed across the assigned scenarios. This was due to the use of student numbers to distribute scenarios and student numbers of participants not being known prior to the study; however, this assignment procedure ensures that our sample of convenience does not impact our results~\cite{druckmanStudentsExperimentalParticipants2011}. Further, our statistical tests are sufficiently robust to account for this uneven distribution~\cite{feirWalshEmpiricalComparisonAnova1974,meyerComparisonExactKruskalWallis2013}, but it is a further reason why we found relatively few significant results. Further, the average grades of participants and non-participants were significantly different, with the participants having a significantly higher average grade, showing that participants self-selected into the study. This may have affected the results, as (more academically capable) students with higher grades may be more likely to be interested in the study and thus more likely to participate. Finally, our participants all self-selected to follow a minor in cybersecurity governance; thus, they have an interest in cybersecurity. While the target population of threat modeling practitioners with a limited technical background might have some understanding of security issues and interest in security~\cite{schieleAcceptability2025}, future research into this target group is needed to confirm this.

Our sample size of 38 is comparatively small; however, it is still a sufficient sample size when compared to the sample sizes of similar studies, which had
102~\cite{schieleAcceptability2025}, 87~\cite{karpatiInvestigatingSecurityThreats2015}, 63~\cite{lallieEmpiricalEvaluationEffectiveness2017}, 63~\cite{opdahlExperimentalComparisonAttack2009}, 49~\cite{broccia2025evaluating}, 43~\cite{freyGoodBadUgly2019}, 42~\cite{kattaComparingTwoTechniques2010}, 28~\cite{labunetsExperimentalComparisonTwo2013}, 28~\cite{grondahlReducingEffortComprehend2011}, and 25~\cite{broccia_assessing_2024}, and 25~\cite{stevensBattleNewYork2018} participants. The small sample size is accounted for in our processing, as the Kruskal-Wallis and Wellek-Welch tests are both robust to small sample sizes. This does explain why we found relatively few significant results, especially in Section~\ref{ssec:results-likert_by_model_within_scenario}, where our sample of 38 was further subdivided.

Finally, the interpretation of our results is colored by the unknown effect of one of our models (AG) not having a model-specific tool. We could have designed the study to require all participants to use the same tool, or no tool at all; however, such a study design would be less reflective of how threat modelling is performed in practice~\cite{verreydt2024threat}: while requiring the participants to use the same tool would result in a more controlled environment, it would not reflect the reality of how threat models are created in industry and thus would potentially be less useful for the field.

\section{Conclusion and Future Work}
\label{sec:conclusion}

Practitioners being able to effectively communicate and analyze threat scenarios is a critical part of the cybersecurity industry. To accomplish these goals, understanding which threat models are best suited for these purposes and under which conditions they are best suited is important. Our study sought to understand the acceptability of three different general graphical threat models in a non-technical population.

We conducted a study with 38 student participants. Participants created AGs, ADTs, and CORAS models of three scenarios according to a Latin square design. Participants then answered perception questions related to the MEM. We processed these data alongside qualitative analysis of the submitted threat models. 

We found that the acceptability of the models did not differ significantly, and that the scenario the model was used on did not affect the acceptability of the model. We also found that the lack of a specific tool for AGs may have impacted the perceived usefulness of AGs. These findings underscore the idea that the choice of threat model is not as important as the fact that a threat model is used. We further find that both ADTs and CORAS are broadly acceptable for a wide range of scenarios, with CORAS showing the most consistent results. These findings can guide practitioners in selecting TM methods in their own contexts.

As mentioned previously, this work would benefit from being performed on a larger sample to both confirm the findings and to improve the statistical power of the tests performed. Additionally, the interesting findings w.r.t. the effect of tooling on acceptability should be further investigated.

\bibliography{bibliography}{}
\bibliographystyle{plain}

% \appendix
% \input{content/questionnaire}

\end{document}